\documentclass[nofootinbib,showkeys]{revtex4}%
\usepackage{etoolbox}
\usepackage[colorlinks=true]{hyperref}
\AtBeginDocument{
	\hypersetup{
		urlcolor=cyan,
		citecolor=magenta,
		linkcolor=blue,
		anchorcolor=green,
	}%
}
\usepackage{amssymb}
\usepackage{amsthm}
\usepackage{amsmath}
\usepackage{extpfeil}
\usepackage{mathtools}
\usepackage{amsfonts}
\usepackage{graphicx}
\usepackage{ulem} 
\usepackage{cancel}
\usepackage[usenames]{xcolor}
\usepackage{epstopdf}
\usepackage{extarrows}
\usepackage{multirow}

\usepackage{tikz}
\usetikzlibrary{matrix,arrows}

\newtheorem{theorem}{Theorem}[section]
\newtheorem{lemma}[theorem]{Lemma}

\newtheorem{corollary}{Corollary}[theorem]

\providecommand{\aln}[1]{\ln_\alpha\left( #1 \right)} %
\providecommand{\sgn}[1]{\mathrm{sgn} \left( #1 \right) }%
\providecommand{\mtn}[1]{{\scriptscriptstyle \!\, #1}}%
\providecommand{\rbr}[1]{\left( #1 \right)}%
\providecommand{\sqbr}[1]{\left[ #1 \right]} %
\definecolor{lred}{rgb}{1,0.3,0}
\definecolor{dgreen}{rgb}{0.0,0.6,0}

\definecolor{amaranth}{rgb}{0.9, 0.17, 0.31}
\definecolor{purple(munsell)}{rgb}{0.62, 0.0, 0.77}
\definecolor{americanrose}{rgb}{1.0, 0.01, 0.24}
\definecolor{palatinateblue}{rgb}{0.15, 0.23, 0.89}
\definecolor{royalblue(web)}{rgb}{0.25, 0.41, 0.88}
\definecolor{hanpurple}{rgb}{0.32, 0.09, 0.98}
\definecolor{beaublue}{rgb}{0.74, 0.83, 0.9}
\definecolor{carminered}{rgb}{1.0, 0.0, 0.22}
\definecolor{brightpink}{rgb}{1.0, 0.0, 0.5}
\definecolor{vividviolet}{rgb}{0.62, 0.0, 1.0}
\hypersetup{ linktoc=all,
	colorlinks, linkcolor={palatinateblue},
	citecolor={brightpink}, urlcolor={amaranth}}

\def\sideremark#1{\ifvmode\leavevmode\fi\vadjust{\vbox to0pt{\vss
			\hbox to 0pt{\hskip\hsize\hskip1em
				\vbox{\hsize2cm\tiny\raggedright\pretolerance10000
					\noindent #1\hfill}\hss}\vbox to8pt{\vfil}\vss}}}%

\makeatletter
\newcommand{\qbar}{\text{\q@bar}}
\newcommand{\q@bar}{%
	\vphantom{$\m@th q$}%
	\ooalign{%
		$\m@th q$\cr
		\hidewidth\kern.15em\smash{\raisebox{-0.7ex}{$\m@th\mathchar'26$}}\hidewidth\cr}%
}
\makeatother

\begin{document}
	
\title[ ]{Multipartite Correlated Majorization Criteria for Finite Discrete Probability Distributions}
\author{Thomas Oikonomou}
\affiliation{College of Business and Management, VinUniversity, Hanoi, Vietnam}
\email{thomas.oikonomou1@gmail.com}
\author{G. Baris Bagci}
\affiliation{Department of Physics, Mersin University, Mersin 33343, Turkey}
\email{gbb0002@hotmail.com }
\author{Charles Casimiro Cavalcante}
\affiliation{Department of Teleinformatics Engineering, Federal University of Ceara, Fortaleza, Brazil}
\email{charles@ufc.br}

%
%

%
	

\begin{abstract}
In this paper we study multipartite and correlated majorization of the finite discrete probability distributions emerging in quantum information theory. We start proving the subadditivity of the R\'{e}nyi and Burg entropies, and we show that the criteria for such a generalized majorization scheme can be provided solely in terms of the R\'{e}nyi and Burg entropies. Surprisingly, the same set of criteria applies both to the correlated and uncorrelated cases. Finally, based on our findings in majorization, we give a proof of the characterization of the R\'{e}nyi and Burg entropies in terms of continuity, symmetry and (sub)additivity.
\end{abstract}

\eid{ }
\date{\today }
\startpage{1}
\endpage{1}
\maketitle

\section{Introduction}

Entropy is the key concept to quantify information. Its applications span the fields such as information theory, thermodynamics, communication, econometric and other physical systems \cite{Cover:2006}. This journal started when Claude Shannon observed that information was linked to probability and proposed a quantity named entropy to measure information or uncertainty \cite{Shannon_1948}. While entropy indeed quantifies information and uncertainty, it is also considered as a measure of disorder. In this context, entropy is linked to another important concept called \textit{majorization} \cite{Marshall}. In fact, when a discrete probability distribution $p$ majorizes another discrete probability distribution $q$, denoted as $p \succ q$, $q$ is said to be more disordered than $p$. The link between entropy and disorder is particularly essential in thermodynamics where majorization is used to determine the thermodynamically allowed state transitions $p \rightarrow q$ in out-of-equilibrium systems without the energy constraints \cite{horodecki, brandao}.

On the other hand, thermodynamic systems require the presence of another system, namely, a heat bath \cite{brandao}. Hence, one needs to generalize the notion of majorization in a multipartite setting such as $p \otimes r \succ q \otimes r$ where $r$ denotes the reservoir probability distribution and $ \otimes$ denotes the Kronecker product. This type of majorization relation is called \textit{trumping} and denoted by $p \succ_T q$ \cite{trump, Klimesh, Turgut}. While the necessary and sufficient condition for majorization $p \succ q$ is the existence of a bistochastic map $\Phi$ such that $q = \Phi (p)$, the relation $p \succ_T q$ is satisfied if and only if the discrete probability distributions $p$ and $q$ satisfy certain relations in terms of the R\'{e}nyi and Burg entropies \cite{renyi, burg} (see Lemma (\ref{Lemma2} in Ref. \cite{brandao}).

However, trumping relations do not take into account the correlations. Motivated by this fact, a new type of trumping relation, named $c$-trumping, has recently been introduced  \cite{MuellerPastena}. The $c$-trumping, denoted by $p \succ_c q$, investigates the transitions of the form $p \otimes(r_1\otimes r_2 \otimes \ldots \otimes r_k) \succ q \otimes r_{1,2,\ldots,k}$ where $ r_{1,2,\ldots,k}$ is a $k$-partite distribution with marginal distributions  $r_1\otimes r_2 \otimes \ldots \otimes r_k$ and $k \geq 3$. M\"uller and Pastena~\cite{MuellerPastena} have shown that the relation $p \succ_c q$ is satisfied if and only if $\text{rank}(p) \leq \text{rank}(q)$ and $H(p) < H(q)$ where $H(p)$ denotes the Shannon entropy and $\text{rank}(p)$ is the number of non-zero entries of the discrete probability distribution $p$.

Extending the majorization such as to include correlations requires an important property to be satisfied by the entropy expression, namely, subadditivity. In fact, this is the main reason why the  criterion for the $c$-trumping includes the Shannon entropy \cite{MuellerPastena}. Subadditivity, and its counter part superadditivity, have been shown to be a powerful tool for both classical and quantum information theory \cite{nielsen}. Its applications span from the areas of channel capacity determination in ``classical'' communication systems to quantum entanglement and measuring the distinguishability of states by concomitant density matrices in quantum information processing. In the classical domain, Anantharam \textit{et al} \cite{Nair:2019} investigated some properties of subadditivity for expressions of mutual information to derive capacity regions, especially for the Gaussian optimality when using such functionals. Also, Liu \textit{et al} investigated the Brascamp-Lieb inequalities (BLI), which has a strong relationship to the subadditivity for certain functionals \cite{Carlen:2009}. These works follow the steps of Geng and Nair \cite{Geng:2014} who showed that subadditivity relations for functionals are sufficient to establish Gaussian optimality and to derive capacity regions. The same approach was considered in several works such as \cite{Lieb:2002,Carlen:1991,Courtade:2014,Zhang:2018,Courtade:2018}. We have also been observing the importance of superadditivity in problems of multicriteria decision aiding (MCDA), where the Choquet integral \cite{Marichal:2000,Grabisch:2010} has been used to provide data aggregation extending the linear strategy to nonlinear cases in which subadditivity and superadditivity behavior can be observed and then redundancy and synergy features can be explored \cite{Duarte:2022}.

It is not accidental then that the conditions for $c$-trumping i.e., majorization relations endowed with correlation and multi-partiteness, require the subadditive Shannon entropy \cite{MuellerPastena}. In this work, we show that the R\'{e}nyi and Burg entropies are also subadditive, and hence a more general criterion for the $c$-trumping can be provided in terms of them instead of the Shannon entropy which is only the $\alpha \rightarrow 1$ limit of the R\'{e}nyi entropy \cite{renyi}. Moreover, we provide a corollary of this result which states that a continuous function having the properties of symmetry, subadditivity and additivity can only be a combination of the R\'{e}nyi and Burg entropies.

The remainder of this paper is organized as follows. In Section \ref{SecII}, we provide the fundamental definitions and properties necessary to state our main results, Theorem \ref{theorem1} and its Corollary. In Section \ref{SecIII}, the subadditivity of the R\'{e}nyi and Burg entropies when considering correlated states are shown. The required lemmas, and the proofs of Theorem \ref{theorem1} and its Corollary are given in Section \ref{SecIV}. Finally, our conclusions are stated in Section \ref{SecCon}.

\section{Definitions and Properties} \label{SecII}

The \textit{R\'{e}nyi entropy} \cite{renyi}, for any probability distribution $p\in \mathcal{I}^+_m$, $\mathcal{I}^+_m:=\{(p_1,\ldots,p_m)\in\mathbb{R}^m\Big|\; p_i>0, \;\sum_{i=1}^{m}p_i=1\}$, is defined as
\begin{eqnarray}\label{renyi_entropy}
H_\alpha(p):=\frac{\mathrm{sgn}(\alpha)}{1-\alpha}\ln\left(\sum_{i=1}^{m}p_i^\alpha\right)\,,
\end{eqnarray}
where $\alpha\in\overline{\mathbb{R}}:=\{x|-\infty\leq x \leq +\infty\}$ denotes the order of the measure, $\ln(x)$ is the natural logarithm, and $\sgn{x}$ is defined as
\begin{eqnarray}\label{signum}
\sgn{x}:=\begin{cases}
+1\,,& x\geq0\\ - 1\,, &x<0
\end{cases}\,.
\end{eqnarray}
%
%
The R\'{e}nyi entropy is  Schur-concave for $\alpha = \{-\infty,0,+\infty\}$, strictly Schur-concave for $\alpha\in(-\infty, 0)\cup(0,\infty)$ \cite{brandao} and additive for every $\alpha\in \mathbb{R}$ \cite{Marshall}. In addition, we have, for $\alpha \to 1$
\begin{eqnarray}\label{shannon_entropy}
H_1(p):=-\sum_{i=1}^{m}p_i \ln(p_i)\,,
\end{eqnarray}
which is the \textit{Shannon entropy}. The \textit{Hartley entropy} \cite{MuellerPastena} reads as
\begin{eqnarray}\label{hartley_entropy}
H_0(p):=\ln \text{rank}(p)\,,	
\end{eqnarray}
while the \textit{Burg entropy} \cite{burg} is defined as
\begin{eqnarray}\label{burg_entropy}
H_\mathrm{Burg}(p):=\frac{1}{m}\sum_{i=1}^{m}\ln(p_i)\,,
\end{eqnarray}
which is strictly Schur-concave \cite{brandao}.

We also define the \textit{deformed $\alpha$-logarithmic} and \textit{$\alpha$-exponential} functions, respectively, as \cite{Tsallis2}
\begin{eqnarray}\label{eq00a}
\aln{x}:=\frac{x^{1-\alpha}-1}{1-\alpha}\,,\qquad
\exp_\alpha(x):=\big[1+(1-\alpha)x\big]^{\frac{1}{1-\alpha}}\,,
\end{eqnarray}
with $1+(1-\alpha)x\geq 0$. The function $\aln{x}$ is concave for $\alpha>0$ and convex for $\alpha<0$. For $\alpha=0$, it is both convex and concave, since $\ln_0(x)=x-1$ is a linear function.
In particular,  the deformed $\alpha$-logarithm satisfies the following:
\begin{subequations}\label{aadd}
	\begin{eqnarray}
	\label{aadd1}
	\aln{xy} &=& \aln{x}+\aln{y}+(1-\alpha)\aln{x}\aln{y}\,,
	\hskip1.0cm (\alpha\in\mathbb{R}\,,\quad x,y>0)\,,\\
	\label{aadd2}
	\aln{x/y}&=& y^{\alpha-1}\big[\aln{x}-\aln{y}\big]\,,
	\hskip3.6cm (\alpha\in\mathbb{R}\,,\quad x,y>0)\,,\\
	\label{aadd3}
	\aln{x}&=&-x^{1-\alpha}\aln{1/x}\,,
	\hskip4.75cm (\alpha\in\mathbb{R}\,,\quad\hskip0.3cm x>0)\,.
	\end{eqnarray}
\end{subequations}
and
\begin{eqnarray}\label{aProp}
\begin{rcases}
\aln{x}\leq x-1\,,& a\geq0\,,\;x>0\\
\aln{x}\geq x-1\,,& a\leq0\,,\;x>0
\end{rcases} \qquad\Rightarrow\qquad
\sgn{\alpha}\aln{x}\leq \sgn{\alpha}(x-1)\,, \quad (\alpha\in\mathbb{R}\,,\;x>0)\,
\end{eqnarray}

The Kronecker product $p\otimes q$ between two probability vectors $p,q\in\mathcal{I}^+_m$ is defined as
\begin{eqnarray}
p\otimes q=
\begin{pmatrix}
p_1q_1 & p_1q_2 & p_1q_3 & \cdots & p_1q_m\\
p_2q_1 & p_2q_2 & p_2q_3 & \cdots & p_2q_m\\
\vdots & \vdots & \vdots & \cdots & \vdots\\
p_m q_1 & p_m q_2 & p_m q_3 & \cdots & p_m q_m
\end{pmatrix}\,.
\end{eqnarray}
Based on these definitions and properties, we may express \textit{R\'{e}nyi divergence} of order $\alpha$ \cite{brandao}, denoted by $D^\mtn{R}_\alpha (p||q)$, as
\begin{subequations}\label{eq01}
	\begin{eqnarray}
	\label{eq01R}
	D^\mtn{R}_\alpha (p||q) &:=&
	\frac{\mathrm{sgn}(\alpha)}{\alpha-1} \ln\rbr{\sum_{i=1}^{m} \rbr{\frac{p_i}{q_i}}^\alpha q_i}
	=
	\mathrm{sgn}(\alpha)\ln
	\left(\exp_{2-\alpha}\Big(\mathrm{sgn}(\alpha)
	D^\mtn{T}_\alpha (p||q) \Big)
	\right)\,,\\
	\label{eq01T}
	D^\mtn{T}_\alpha (p||q) &:=&
	\frac{\mathrm{sgn}(\alpha)}{\alpha-1}\sqbr{\sum_{i=1}^{m}\rbr{\frac{p_i}{q_i}}^\alpha q_i-1}
	=-\mathrm{sgn}(\alpha)\sum_{i=1}^{m}p_i\ln_{\alpha}\rbr{\frac{q_i}{p_i}}\,.
	\end{eqnarray}
\end{subequations}
The auxiliary expression $D^\mtn{T}_\alpha (p||q)$ is known as the \textit{Tsallis divergence} \cite{Tsallis2,Oik1}.

When one has product states  as the arguments with the probability vectors $p,q\in\mathcal{I}^+_m $ and $r,s\in\mathcal{I}^+_n$, R\'{e}nyi divergence, with $D^\mtn{T}_\alpha (p\otimes r||q\otimes s) = -\sgn{\alpha}\sum_{ij}p_ir_j\aln{q_is_j/p_ir_j}$, is additive
%
\begin{eqnarray}
\label{TenProdTsDivR}
D_\alpha^\mtn{R}\big(p\otimes r||q\otimes s\big)&=& D_\alpha^\mtn{R}\big(p||q\big) + D_\alpha^\mtn{R}\big(r|| s\big)
\end{eqnarray}
%
for $\alpha\in\mathbb{R}$ which can be shown by using Eq.~(\ref{eq01}).


Invoking the Jensen's inequality for convex or concave functions, i.e., $\mathbb{E}\{f(X)\}\geq f(\mathbb{E}\{X\})$ or $\mathbb{E}\{f(X)\}\leq f(\mathbb{E}\{X\})$, respectively, one can prove the nonnegativity of the Tsallis divergence and accordingly via Eq. (\ref{eq01R}) of the R\'{e}nyi divergence we have
\begin{eqnarray}\label{TsDivConvex}
D^\mtn{T}_\alpha (p||q)
\geq 0\,, \qquad
D^\mtn{R}_\alpha (p||q)
\geq 0
\,,\qquad (\forall\alpha\in\mathbb{R})\,.
\end{eqnarray}

We also note that
\begin{eqnarray}\label{renyi_uniform}
D^\mtn{R}_\alpha (p||\eta) = \sgn\alpha\ln (m)-H_\alpha(p) 	
\end{eqnarray}
where $\eta=\lbrace \frac{1}{m},\frac{1}{m},...,\frac{1}{m}  \rbrace$ is the uniform probability distribution.

The Burg divergence then reads \cite{Gorban,Burg}
\begin{eqnarray}\label{BurgDiv}
D_\mathrm{Burg}(p||q):=-\sum_{i=1}^{m}q_i\ln\left(\frac{p_i}{q_i}\right)
\end{eqnarray}
with
\begin{eqnarray}\label{BurgDivEnt}
D_\mathrm{Burg}(p||\eta):=-\ln(m) - H_\mathrm{Burg}(p)\,.
\end{eqnarray}

Having provided the fundamental definitions and properties, we can state our main result:
\begin{theorem}\label{theorem1}
	Let $q,p\in\mathcal{I}^+_m$ be probability distributions with 	$p^{\downarrow}\neq q^{\downarrow}$ where $p^{\downarrow}$ ($q^{\downarrow}$) denotes the reordering of the entries of the probability distribution $p$ ($q$) in descending order. Then, there exists $k\in\mathbb{N}_0$ and a $ k $-partite probability distribution $r_{1,2,\ldots,k}$,  with marginals $r_1, r_2, \ldots, r_k$, such that
	\begin{eqnarray}\label{eq01B}
	p\otimes(r_1\otimes\cdots\otimes r_k) \succ q\otimes r_{1,\ldots,k}
	\end{eqnarray}
	if and only if $\text{rank}(p) \leq \text{rank}(q)$, $H_{\alpha}(p) < H_{\alpha}(q)$ for $\alpha\in(0,\infty)$ and $H_\mathrm{Burg}(p)<H_\mathrm{Burg}(q)$. One can always set $k=2$.
\end{theorem}
Let us now define $\mathcal{I}^{+}:= \bigcup_{n \in \mathbb{N}}\mathcal{I}_n^{+}$. Then, the immediate corollary of Theorem \ref{theorem1} can be given as:
\begin{corollary}\label{last3}	A continuous function $S : \mathcal{I}^+\to\mathbb{R}$ satisfies the following three properties
	\begin{itemize}
		\item[(i)] \textbf{symmetry:} if $p, q \in \mathcal{I}^+$ are such that $p_i = q_{\pi(i)}$ for some permutation $\pi$ and all $ i $, then $ S(p) = S(q) $;
		\item[(ii)] \textbf{subadditivity:} $ S(p_{AB}) \leq S(p_A \otimes p_B) $ for every bipartite
		probability distribution $p_{AB} \in \mathcal{I}^+$ with marginals $ p_A $ 	and $ p_B $;
		\item[(iii)] \textbf{additivity:} $S(p_A\otimes p_B) = S(p_A)+S(p_B)$ for all $ p_A, p_B \in \mathcal{I}^+$
	\end{itemize}
	if and only if it is of the form
	\begin{eqnarray}\label{last2}	
	S(p)=d_1\cdot H_{\alpha}(p)+d_2\cdot  H_\mathrm{Burg}(p)+c_{n} \qquad \forall p\in\mathcal{I}^+\,,\quad n\in\mathbb{N}\,,\quad \alpha\in(0,+\infty)\,,
	\end{eqnarray}
	where $H_\alpha(p)$ is the R\'{e}nyi entropy, $H_\mathrm{Burg}(p)$ is the Burg entropy, $ c_{n}  \geq 0$ ($ c_{n}  \in \mathbb{R}$) is some dimension dependent constant, and $d_{1}, d_{2} \geq 0$.
\end{corollary}

A pivotal component for the proof of the preceding Theorem and its Corollary is the subadditivity of R\'{e}nyi and Burg entropies. Hence, next section is devoted to the proof of the subadditivity of these entropies.

\section{Subadditivity of R\'{e}nyi and Burg entropies for correlated states}\label{SecIII}

We begin by proving the superadditivity of the related divergences \cite{Perez}, first focusing on the Tsallis divergence \cite{tsallis_gen} .
\begin{lemma}\label{tsallis_div_superadd}
	The Tsallis divergence for correlated states is $\alpha$-superadditive, it means that
	\begin{eqnarray}\label{TsDivaSA}
	D^\mtn{T}_\alpha(p_\mtn{AB}||q_\mtn{A}\otimes q_\mtn{B}) \geq D^\mtn{T}_\alpha(p_\mtn{A}\otimes p_\mtn{B}||q_\mtn{A}\otimes q_\mtn{B})
	=
	D^\mtn{T}_\alpha(p_\mtn{A}||q_\mtn{A})
	+
	D^\mtn{T}_\alpha( p_\mtn{B}|| q_\mtn{B})
	+
	\sgn{\alpha}(\alpha-1)
	D^\mtn{T}_\alpha(p_\mtn{A}||q_\mtn{A})
	D^\mtn{T}_\alpha( p_\mtn{B}|| q_\mtn{B})
	\end{eqnarray}
	for $\alpha\in\mathbb{R}$.
\end{lemma}

\begin{proof}
	Consider the discrete probability vectors $q_\mtn{A},p_\mtn{A}\in\mathcal{I}_n^+$ and $q_\mtn{B},p_\mtn{B}\in\mathcal{I}_m^+$.
	For simplicity, we will use the index $i$ ($j$) to describe the probability vector components in system $A$ ($B$), i.e., $q_\mtn{A}=\{q_i\}_{i=1}^{n}$,
	$q_\mtn{B}=\{q_j\}_{j=1}^{m}$,
	$p_\mtn{A}=\{p_i\}_{i=1}^{n}$,
	$p_\mtn{B}=\{p_j\}_{j=1}^{m}$.
	We also have $p_\mtn{AB}\in\mathcal{I}_{nm}^+$ with $p_i=\sum_j p_{ij}$ while $p_{ij}\neq p_ip_j$, in general. Before proceeding further,  we determine the sign of the following expression for $\alpha\neq 1$
	\begin{eqnarray}\label{TheA}
	A:=
	-\sgn{\alpha}\alpha\sum_{ij} (p_{ij}-p_ip_j) \aln{\frac{q_i q_j}{p_i p_j}}
	=
	\sgn{\alpha}
	\frac{\alpha}{\alpha-1}
	\sum_{ij} (p_{ij}-p_ip_j) \rbr{\frac{q_i q_j}{p_i p_j}}^{1-\alpha}\,.
	\end{eqnarray}
	One can easily verify that $A=0$ for $\alpha \rightarrow 1$, since  $\sum_{ij}(p_{ij}-p_ip_j)f(x_i)=0$,
	for any real continuous function $f(x_i)$, where the marginal distribution $p_i=\sum_{j}p_{ij}$ is taken into account (this result also holds when one switches the indices).
	For simplicity, we denote $\mathcal{P}_{ij}:=\sgn{\alpha}\frac{\alpha}{\alpha-1}(p_{ij}-p_ip_j)$, $x_{i}=(q_i/p_i)^{1-\alpha}>0$ and $y_{j}=(q_j/p_j)^{1-\alpha}>0$. Taking into account that  $\sum_{ij}\mathcal{P}_{ij}
	x_i e^{b x_i}\sim \sum_{ij}(p_{ij}-p_ip_j)f(x_i)=0$, where $f(x_i):=x_ie^{b x_i}$ for any $b\in\mathbb{R}$, the term $A$ can be rewritten as
	\begin{eqnarray*}\label{TheA2}
		A
		=
		\sum_{ij}\mathcal{P}_{ij} x_i  y_j
		=
		\sum_{ij}\mathcal{P}_{ij} x_i y_j
		+
		\sum_{ij}\mathcal{P}_{ij}
		x_i e^{b x_i}
		=
		\sum_{ij}\mathcal{P}_{ij} x_i  \left(y_j+e^{bx_i}\right)\,.
	\end{eqnarray*}
	Then, if we split the matrix $\mathcal{P}_{ij}$ into two matrices consisting of nonnegative and negative components, respectively,  as $\mathcal{P}_{ij}=\mathcal{P}_{ij}^{+} + \mathcal{P}_{ij}^{-}$, we have
	\begin{eqnarray*}
		A = \sum_{ij}\mathcal{P}^+_{ij}x_i(y_j+e^{b x_i}) + \sum_{ij}\mathcal{P}^-_{ij}x_i(y_j+e^{b x_i}).\
	\end{eqnarray*}
	For $\epsilon>0$ there is $b$  such that $(y_j)_{\max}+e^{b x_i} < e^{(b+\epsilon) x_i} \;\Rightarrow\; y_j+e^{b x_i} < e^{(b+\epsilon) x_i}$. Then, by multiplying the former inequality by $\mathcal{P}^{-}_{ij}x_i$, summing over $i,j$, and adding the term $\sum_{ij}\mathcal{P}^{+}x_i(y_j +e^{b x_i})$ on both sides, the l.h.s. becomes equal to $A$, and we accordingly have
	\begin{eqnarray*}
		A>\sum_{ij}\mathcal{P}_{ij}^{+}x_i(y_j +e^{b x_i}) + 
		\sum_{ij}\mathcal{P}_{ij}^{-}x_ie^{(b+\epsilon) x_i}
		=
		\sum_{ij}\mathcal{P}_{ij}^{+}x_iy_j 
		+ \sum_{ij} \left( 
		\mathcal{P}_{ij}^{+}
		+
		\mathcal{P}_{ij}^{-}e^{\epsilon x_i} \right) x_i e^{b x_i}
	\end{eqnarray*}
	Note that one can arbitrarily choose a sufficiently large $b$  such that $\epsilon\to0$. Then, the former inequality becomes
	\begin{eqnarray*}
		A
		>
		\sum_{ij}\mathcal{P}_{ij}^{+}x_iy_j 
		+ \sum_{ij} \left( 
		\mathcal{P}_{ij}^{+}
		+
		\mathcal{P}_{ij}^{-}e^{\epsilon x_i} \right) x_i e^{b x_i}
		\approx
		\sum_{ij}\mathcal{P}^{+}x_iy_j 
		+ \sum_{ij} \left( 
		\mathcal{P}_{ij}^{+}
		+
		\mathcal{P}_{ij}^{-}\right) x_i e^{b x_i}
	\end{eqnarray*}
	The last term with $\mathcal{P}_{ij}=\mathcal{P}_{ij}^{+} + \mathcal{P}_{ij}^{-}$ as explained above is equal to zero, so that
	\begin{eqnarray}\label{wow}
	A
	>
	\sum_{ij}\mathcal{P}_{ij}^{+}x_iy_j 
	+ \sum_{ij} \left( 
	\mathcal{P}_{ij}^{+}
	+
	\mathcal{P}_{ij}^{-}e^{\epsilon x_i} \right) x_i e^{b x_i}
	\approx
	\sum_{ij}\mathcal{P}^{+}x_iy_j 
	>0\,.
	\end{eqnarray}	
	Herewith, we determined the sign of $A$ to be positive.

	Turning now to prove $\alpha$-superadditivity of the Tsallis divergence for correlated states i.e. Eq. (\ref{TsDivaSA}), we have

	\begin{subequations}\label{SupAdd}
		\begin{eqnarray}
		\label{SupAdd1}
		D^\mtn{T}_\alpha(p_\mtn{AB}||q_\mtn{A}\otimes q_\mtn{B})
		&=&
		-\sgn{\alpha}\sum_{ij}p_{ij}\aln{\frac{q_iq_j}{p_ip_j}}
		+
		\sgn{\alpha}\sum_{ij}p_{ij}\sqbr{
			\aln{ \frac{ q_iq_j}{p_ip_j}}
			-
			\aln{\frac{q_iq_j}{p_{ij}}} }\\
		\label{SupAdd2}
		&{=}&
		-\sgn{\alpha}\sum_{ij}p_{ij}\aln{\frac{q_iq_j}{p_ip_j}}
		+
		\sgn{\alpha}	
		\sum_{ij}p_{ij} \rbr{\frac{q_iq_j}{p_{ij}}}^{1-\alpha}
		\aln{\frac{p_{ij}}{p_ip_j}}\\
		\label{SupAdd3}
		&{=}&
		-\sgn{\alpha}\sum_{ij}p_{ij} \aln{ \frac{ q_iq_j}{p_ip_j } }
		-\sgn{\alpha}
		\sum_{ij}p_{ij}
		\rbr{ \frac{q_iq_j}{p_{ij}}}^{1-\alpha}
		\rbr{\frac{p_{ij}}{p_ip_j}}^{1-\alpha}
		\aln{\frac{p_ip_j}{p_{ij}}}\\
		\label{SupAdd4}
		&=&
		-\sgn{\alpha}\sum_{ij}p_{ij} \aln{ \frac{ q_iq_j}{p_ip_j}}
		-\sgn{\alpha}
		\sum_{ij}p_{ij}
		\rbr{ \frac{ q_iq_j}{p_ip_j}}^{1-\alpha}
		\aln{\frac{p_ip_j}{p_{ij}}}\\
		\label{SupAdd5}
		&{\geq}&
		-\sgn{\alpha}\sum_{ij}p_{ij}
		\aln{ \frac{ q_iq_j}{p_ip_j}}
		+\sgn{\alpha}
		\sum_{ij}p_{ij}
		\rbr{ \frac{ q_iq_j}{p_ip_j}}^{1-\alpha}
		\rbr{1-\frac{p_ip_j}{p_{ij}}}\\
		\label{SupAdd6}
		&=&
		-\sgn{\alpha}\sum_{ij}p_{ij} \aln{ \frac{ q_iq_j}{p_ip_j}}
		+\sgn{\alpha}
		\sum_{ij}p_{ij}
		\rbr{ \frac{ q_iq_j}{p_ip_j}}^{1-\alpha}
		-
		\sum_{ij}p_{i}p_j
		\rbr{ \frac{ q_iq_j}{p_ip_j}}^{1-\alpha}\\
		\nonumber
		&=&
		-\sgn{\alpha}\sum_{ij}p_{ij}
		\aln{ \frac{ q_iq_j}{p_ip_j}}
		+
		(1-\alpha)\sgn{\alpha}\sum_{ij}p_{ij}
		\aln{ \frac{ q_iq_j}{p_ip_j}}\\
		\label{SupAdd7}
		&&\hskip4.0cm
		-
		(1-\alpha)\sgn{\alpha}
		\sum_{ij}p_{i}p_j
		\aln{ \frac{ q_iq_j}{p_ip_j}}\\
		\label{SupAdd8}
		&=&
		-\sgn{\alpha}\sum_{ij}p_{i}p_j
		\aln{ \frac{ q_iq_j}{p_ip_j}}
		+
		\underbrace{\left[- \sgn{\alpha}\alpha
			\sum_{ij}(p_{ij}-p_ip_j)
			\aln{ \frac{ q_iq_j}{p_ip_j}} \right] }_{A}\\
		\label{SupAdd9}
		&\geq&
		-\sgn{\alpha}	\sum_{ij}p_{i}p_j
		\aln{\frac{q_iq_j}{p_ip_j}}=D^\mtn{T}_\alpha(p_\mtn{A}\otimes p_\mtn{B}||q_\mtn{A}\otimes q_\mtn{B})\,
		\end{eqnarray}
	\end{subequations}
	for all $\alpha\in\mathbb{R}$.  Note that we have used Eqs. (\ref{aadd2}), (\ref{aadd3}), (\ref{aProp}) and (\ref{wow}) in Eqs. (\ref{SupAdd2}),  (\ref{SupAdd3}), (\ref{SupAdd5}) and (\ref{SupAdd9}), respectively.

\end{proof}

\begin{lemma}\label{renyi_div_superadd}
	The R\'{e}nyi divergence for correlated states is superadditive for $\alpha\in\mathbb{R}$,
	\begin{eqnarray}\label{renyiDivaSA}
	D^\mtn{R}_\alpha(p_\mtn{AB}||q_\mtn{A}\otimes q_\mtn{B}) \geq D^\mtn{R}_\alpha(p_\mtn{A}\otimes p_\mtn{B}||q_\mtn{A}\otimes q_\mtn{B})\,.
	\end{eqnarray}
\end{lemma}

\begin{proof}
	
	\begin{subequations}
		\begin{eqnarray}
		D_\alpha^\mtn{R}(p_\mtn{AB}||q_\mtn{A}\otimes q_\mtn{B})
		&=&
		D_\alpha^\mtn{R}(p_\mtn{A}\otimes p_\mtn{B}||q_\mtn{A}\otimes q_\mtn{B})\\
		&+&
		\sgn{\alpha}
		\ln\left\{ \exp_{2-\alpha}\Big(\sgn{\alpha}  D^\mtn{T}_\alpha (p_\mtn{AB}||q_\mtn{A}\otimes q_\mtn{B}) \Big) \right\}\\
		&-&
		\sgn{\alpha}
		\ln\left\{ \exp_{2-\alpha}\Big(\sgn{\alpha}  D^\mtn{T}_\alpha (p_\mtn{A}\otimes p_\mtn{B}||q_\mtn{A}\otimes q_\mtn{B}) \Big) \right\}\\
		\label{eq13de}
		&=&
		D_\alpha^\mtn{R}(p_\mtn{A}\otimes p_\mtn{B}||q_\mtn{A}\otimes q_\mtn{B})
		+
		\sgn{\alpha}
		\ln\left\{ \frac{\exp_{2-\alpha} \Big(\sgn{\alpha}  D^\mtn{T}_\alpha (p_\mtn{AB}||q_\mtn{A}\otimes q_\mtn{B}) \Big)}{\exp_{2-\alpha}\Big(\sgn{\alpha}  D^\mtn{T}_\alpha (p_\mtn{A}\otimes p_\mtn{B}||q_\mtn{A}\otimes q_\mtn{B}) \Big)} \right\}
		\end{eqnarray}
	\end{subequations}
	The last logarithmic argument in Eq. (\ref{eq13de}) is equal to unity for $\alpha=0$, greater than unity for $\alpha> 0$, and less than unity for $\alpha<0$ because of $D^\mtn{T}_\alpha(p_\mtn{AB}||q_\mtn{A}\otimes q_\mtn{B})\geq D^\mtn{T}_\alpha(p_\mtn{A}\otimes p_\mtn{B}||q_\mtn{A}\otimes q_\mtn{B})$, so that
	\begin{eqnarray}
	D_\alpha^\mtn{R}(p_\mtn{AB}||q_\mtn{A}\otimes q_\mtn{B})
	&\geq&
	D_\alpha^\mtn{R}(p_\mtn{A}\otimes p_\mtn{B}||q_\mtn{A}\otimes q_\mtn{B})
	\end{eqnarray}
	for $\alpha\in\mathbb{R}$.
	
\end{proof}

\begin{lemma}\label{renyi_ent_subadd}
	For every bipartite probability distribution $p_\mtn{AB}\in\mathcal{I}^+_{nm}$, the R\'{e}nyi entropy is subadditive
	\begin{eqnarray}\label{renyisub1}
	H_\alpha(p_\mtn{AB})  \leq   H_\alpha(p_\mtn{A}\otimes p_\mtn{B})
	\end{eqnarray}
	for $\alpha\in\mathbb{R}$ where $p_\mtn{A}$ and $p_\mtn{B}$ are marginals.
	
\end{lemma}

\begin{proof}
	
	We begin with the superadditivity of the R\'{e}nyi divergence for correlated states, Eq. (\ref{renyiDivaSA}), and set $q_\mtn{A}\otimes q_\mtn{B}$ to be the uniform distribution $\eta$. Then, we obtain
	\begin{eqnarray}\label{renyisub2}
	D^\mtn{R}_\alpha(p_\mtn{AB}||\eta) \geq D^\mtn{R}_\alpha(p_\mtn{A}\otimes p_\mtn{B}||\eta)
	\end{eqnarray}	
	which, using Eq. (\ref{renyi_uniform}), yields
	\begin{eqnarray}\label{renyisub3}
	H_\alpha(p_\mtn{AB}) \leq H_\alpha(p_\mtn{A}\otimes p_\mtn{B})\,.
	\end{eqnarray}

\end{proof}
In Appendix \ref{FirstAppendix}, we prove that R\'{e}nyi entropy does not maintain the subadditivity property for $\alpha=\{-\infty,+\infty\}$.

\begin{lemma}\label{Burg_div_superadd}
	The Burg divergence for correlated states is superadditive,
	\begin{eqnarray}\label{BurgDivaSA}
	D_\mathrm{Burg}(p_\mtn{AB}||q_\mtn{A}\otimes q_\mtn{B}) \geq D_\mathrm{Burg}(p_\mtn{A}\otimes p_\mtn{B}||q_\mtn{A}\otimes q_\mtn{B})\,.
	\end{eqnarray}
\end{lemma}

\begin{proof}
	
	We again consider the probability vectors
	$q_\mtn{A}=\{q_i\}_{i=1}^{n}$,
	$q_\mtn{B}=\{q_j\}_{i=1}^{m}$,
	$p_\mtn{A}=\{p_i\}_{i=1}^{n}$,
	$p_\mtn{B}=\{p_j\}_{j=1}^{m}$.
	We also have $p_\mtn{AB}=\{p_{ij}\}_{i=1,j=1}^{n,m}$ with $p_i=\sum_j p_{ij}$ and $p_j=\sum_i p_{ij}$ while $p_{ij}\neq p_ip_j$ in general. Then, the Burg divergence in Eq. (\ref{BurgDiv}) yields
	\begin{subequations}\label{SupAddb}
		\begin{eqnarray}
		\label{SupAdd1b}
		D_\mathrm{Burg}(p_\mtn{AB}||q_\mtn{A}\otimes q_\mtn{B})
		&=&
		\sum_{ij}q_iq_j \ln\rbr{\frac{q_iq_j}{p_ip_j}}
		-
		\sum_{ij} q_iq_j \sqbr{
			\ln\rbr{ \frac{ q_iq_j}{p_ip_j}}
			-
			\ln\rbr{\frac{q_iq_j}{p_{ij}}} }\\
		\label{SupAdd2b}
		&{=}&
		\sum_{ij}q_iq_j \ln\rbr{\frac{q_iq_j}{p_ip_j}}
		-
		\sum_{ij} q_iq_j
		\ln\rbr{\frac{p_{ij}}{p_ip_j}}\\
		\label{SupAdd3b}
		&\geq &
		\sum_{ij}q_iq_j \ln\rbr{\frac{q_iq_j}{p_ip_j}}
		+
		\sum_{ij}q_iq_j \rbr{1-\frac{p_{ij}}{p_ip_j}}\\
		\label{SupAdd6b}
		&=&
		\sum_{ij}q_iq_j \ln\rbr{\frac{q_iq_j}{p_ip_j}}
		+
		\underbrace{ \sum_{ij} \frac{q_iq_j}{p_ip_j}(p_ip_j-p_{ij})}_{\tilde{A}\geq 0}\\
		\label{SupAdd9b}
		&\geq&
		\sum_{ij}q_iq_j \ln\rbr{\frac{q_iq_j}{p_ip_j}}
		=D_\mathrm{Burg}(p_\mtn{A}\otimes p_\mtn{B}||q_\mtn{A}\otimes q_\mtn{B})
		=D_\mathrm{Burg}(p_\mtn{A}||q_\mtn{A})
		+
		D_\mathrm{Burg}(p_\mtn{B}|| q_\mtn{B}).
		\end{eqnarray}
	\end{subequations}
	To prove $\tilde{A}\geq 0$ in Eq. (\ref{SupAdd6b}), we define $\mathcal{P}_{ij}:=p_ip_j-p_{ij}$, $x_{i}=(q_i/p_i)>0$ and $y_{j}=(q_j/p_j)>0$ so that the term $\tilde{A}$ becomes $\sum_{ij}\mathcal{P}_{ij}
	x_i (y_j+e^{bx_i} )$. Then, we follow the same steps as in Lemma \ref{tsallis_div_superadd}, up to Eq. (\ref{wow}) to prove $\tilde{A}\geq 0$.
\end{proof}

\begin{lemma}\label{renyi_ent_subadd_A}
	For every bipartite probability distribution $p_\mtn{AB}\in\mathcal{I}^+_{nm}$, the Burg entropy is subadditive
	\begin{eqnarray}\label{renyisub1b}
	H_\mathrm{Burg} (p_\mtn{AB})  \leq   H_\mathrm{Burg} (p_\mtn{A}\otimes p_\mtn{B}) \,,
	\end{eqnarray}
	where $p_\mtn{A}$ and $p_\mtn{B}$ are marginals.
	
\end{lemma}

\begin{proof}
	We begin with the superadditivity of the Burg divergence for correlated states, Eq. (\ref{BurgDivaSA}), and set $q_\mtn{A}\otimes q_\mtn{B}$ to be the uniform distribution $\eta$. Then, we obtain
	\begin{eqnarray}\label{Burgsub2}
	D_\mathrm{Burg}(p_\mtn{AB}||\eta) \geq D_\mathrm{Burg}(p_\mtn{A}\otimes p_\mtn{B}||\eta)
	\end{eqnarray}	
	which, using Eq. (\ref{BurgDivEnt}), yields
	\begin{eqnarray}\label{Burgsub3}
	H_\mathrm{Burg}(p_\mtn{AB}) \leq H_\mathrm{Burg}(p_\mtn{A}\otimes p_\mtn{B})\,.
	\end{eqnarray}
\end{proof}

\section{Proof of Theorem \ref{theorem1} and Corollary \ref{last3}}\label{SecIV}

\begin{lemma}\label{Lemma2}  (Trumping \cite{brandao}): Let $p, q \in\mathbb{R}^n$ be probability distributions such that $p^{\downarrow}\neq q^{\downarrow}$, and such that at least one of them has full rank. Then $p \succ_T q$ if and only if
	\begin{eqnarray*}
		H_\alpha(p)&<&H_\alpha(q)\quad \forall \alpha\in\mathbb{R}\backslash\{0\}\,, \; \text{and}\\
		H_\mathrm{Burg}(p)&<&H_\mathrm{Burg}(q)\,.
	\end{eqnarray*}
\end{lemma}
This Lemma states that the R\'{e}nyi and Burg entropies form a complete set of monotones for the trumping relation.\\


For our purpose,  we will consider a particular family of bipartite probability distrivutions.  For any given probability distribution $q_\mtn{A}\in\mathcal{I}_m^+$ of full rank on system $A$, we define the joint probability distribution on $AB$ as \cite{mueller2018}
\begin{eqnarray}\label{jointAB}
q_\mtn{AB}&=&
\begin{pmatrix}
\delta & \delta /n^2 \; \dots \; \delta /n^2 & (q_1-2\delta)/n \;\dots\; (q_1-2\delta)/n\\
\delta & \delta /n^2 \; \dots \; \delta /n^2 & (q_2-2\delta)/n \;\dots\; (q_2-2\delta)/n\\
\vdots & \vdots \hskip1.3cm \vdots & \vdots \hskip2.3cm \vdots\\
\delta & \underbrace{\delta /n^2 \;\dots \; \delta /n^2}_{n^2} & \underbrace{(q_m-2\delta)/n \;\dots\; (q_m-2\delta)/n}_{n}
\end{pmatrix}\,,
\end{eqnarray}
where $n\in\mathbb{N}$ and $0<\delta<\frac{1}{2}\min_i q_i$, of the dimension $m\times(n^2+n+1)$. Its marginal probability distribution on $B$ is
\begin{eqnarray}\label{marginalB}
q_\mtn{B}&=&\left(m \delta, \underbrace{\frac{m\delta}{n^2},\ldots, \frac{m\delta}{n^2}}_{n^2},\underbrace{\frac{1-2m\delta}{n},\ldots, \frac{1-2m\delta}{n}}_{n}\right)
\end{eqnarray}
For this set of distributions we have the following.\\

\begin{lemma}\label{lemma5}
	Let $p,q\in\mathcal{I}_m^+$ be probability distributions with full rank such that $H_1(p)<H_1(q)$. Then, $\delta >0$ with $\delta<\frac{1}{2}\min_i q_i$ and $N\in\mathbb{N}$ such that for $q_\mtn{AB}$ as defined in Eq. (\ref{jointAB}) satisfies
	\begin{eqnarray}
	p_\mtn{A}\otimes q_\mtn{B} \succ_T q_\mtn{AB}\qquad \text{for all}\quad n\geq N\,.
	\end{eqnarray}
\end{lemma}

\begin{proof}
	For $\alpha\in\mathbb{R}$, define the entropy difference
	\begin{eqnarray}\label{deltadef}
	\Delta_n^{(\alpha)}&:=& H_\alpha(q_\mtn{AB})-H_\alpha(p_\mtn{A})-H_\alpha(q_\mtn{B})
	\end{eqnarray}
	Substituting the distributions in Eqs. (\ref{jointAB})-(\ref{marginalB}) into Eq. (\ref{deltadef}), we have
	\begin{eqnarray}
	\Delta_n^{(\alpha)}=
	-H_\alpha(p) + \frac{\sgn{\alpha}}{1-\alpha} \ln\left(\frac{m\delta^\alpha+m\delta^\alpha n^{2(1-\alpha)} +n^{1-\alpha}\sum_{i=1}^m (q_i-2\delta)^\alpha}{(m\delta)^\alpha(1+n^{2(1-\alpha)}) + (1-2m\delta)^\alpha n^{1-\alpha}}\right)\,.
	\end{eqnarray}
	One can prove that this quantity is indeed positive for $\alpha\in\mathbb{R}\backslash\{0\}$ \cite{mueller2018}. Note that the value $\alpha=0$ is included in \cite{mueller2018}, which is a  typo \cite{note}. Similarly, for the Burg entropy we define
	\begin{eqnarray}\label{deltadef2}
	\nonumber
	\Delta_n^\mathrm{Burg}&:=& H_\mathrm{Burg}(q_\mtn{AB}) - H_\mathrm{Burg}(p_\mtn{A}) - H_\mathrm{Burg}(q_\mtn{B})\\
	&=& -H_\mathrm{Burg}(p) + \frac{1}{n^2+n+1}
	\left[
	\frac{n}{m}\sum_{i=1}^{m}\ln(q_i-2\delta) - (n^2+1)\ln(m) - n\ln(1-2m\delta)\right]\,,
	\end{eqnarray}
	which is shown to be positive as well \cite{mueller2018}. Then, from Lemma \ref{Lemma2} follows $p_\mtn{A}\otimes q_\mtn{B}\succ_T q_\mtn{AB}$.
\end{proof}

\begin{lemma}\label{lemma7}
	Let $p,q\in\mathcal{I}_m$ be probability distributions such that $q$ has full rank. If $H_\alpha(p)<H_\alpha(q)$ with $\alpha > 0$ and $H^\mathrm{Burg}(p) < H^\mathrm{Burg}(q) $, then there exists $k\in\mathbb{N}$ with $k\geq 3$ and a $k$-partite distribution $r_{1,2,\ldots,k}$ such that
	\begin{eqnarray}
	p\otimes(r_1\otimes r_2\otimes \cdots\otimes r_k) \succ q\otimes r_{1,2,\ldots,k}
	\end{eqnarray}
\end{lemma}

\begin{proof}
	Considering the two distributions $p_\mtn{A}\otimes q_\mtn{A}$ and $q_\mtn{AB}$, Lemma \ref{lemma5} warrants that there exists a system $C$ such that
	\begin{eqnarray*}
		(p_\mtn{A}\otimes q_\mtn{B})\otimes q_\mtn{C} \succ_T q_\mtn{ABC}\,.
	\end{eqnarray*}
	Then, trumping definition and majorization property lead us to the existence of $k=3$ subsystems in addition to system $A$ \cite{MuellerPastena}.
\end{proof}

Having Lemmas \ref{lemma5} and \ref{lemma7} at our disposal, we are ready to prove Theorem \ref{theorem1}.

\begin{proof}
	Assuming the existence of the auxiliary distribution $r_{1,2,\ldots,k}$. Then, the additivity, subadditivity and strict Schur concavity of the R\'{e}nyi entropies for $\alpha>0$ yields
	\begin{eqnarray*}
		H_\alpha(p)+\sum_{i=1}^{k}H_\alpha(r_i) < H_\alpha(q)+H_\alpha(r_{1,2,\ldots,k}) \leq H_\alpha(q) + \sum_{i=1}^{k}H_\alpha(r_i)\,.
	\end{eqnarray*}

	To prove the converse direction, we consider the not equal up to permutation distributions $p,q\in\mathcal{I}_m$ with $\text{rank}(p)\leq \text{rank}(q)=:\ell$, $p^{\downarrow}=p$ and $q^{\downarrow}=q$, and  $H_\alpha(p) < H_\alpha(q)$. Apparently we have $m\geq \ell$. We also introduce the operator $\oplus$, which creates out of two vectors $X\in\mathcal{I}_m$ and $Y\in\mathcal{I}_\ell$ a new vector $X\oplus Y=Z\in\mathcal{I}_{m+\ell}=\{X_1,\ldots,X_m,Y_1,\ldots,Y_\ell\}$. 
	Then, we can rewrite $q$ as $q=\tilde{q}\oplus 0_{m-\ell}$, where $\tilde{q}=(q_1,\ldots,q_\ell)\in\mathcal{I}_{\ell}$ is of full rank, and $0_{m-\ell}=(0,\ldots,0)\in\mathcal{I}_{m-\ell}$ is the zero vector of dimension $m-\ell$. Respectively, for $p$ we can write $p=\tilde{p}\oplus 0_{m-\ell}$, where $\tilde{p}\in\mathcal{I}_{\ell}$ may not have full rank. Then, Eq. (\ref{eq01B}) is equivalent to
	\begin{eqnarray*}
		\tilde{p}\otimes (r_1\otimes r_2\otimes\ldots\otimes r_k) \succ \tilde{q}\otimes r_{1,2,\ldots,k}\,.
	\end{eqnarray*}
	Since $H_\alpha(\tilde{p})=H_\alpha(p) < H_\alpha(q)=H_\alpha(\tilde{q})$ and $\tilde{q}$ has full rank, Lemma \ref{lemma7} warrants the existence of the distribution $r_{1,2,\ldots,k}$ with $k\geq 3$.
	
	The former discussion can be extended to the case $k=2$ too, by showing additionally that for any $\varepsilon>0$ the R\'{e}nyi mutual information $
	I_\alpha(A:B)\equiv D_\alpha(q_\mtn{AB}||q_\mtn{A}\otimes q_\mtn{B})$ satisfies the relation $I_\alpha(A:B)<\varepsilon$ \cite{mueller2018}.
	For this, we substitute Eqs. (\ref{jointAB})-(\ref{marginalB}) into the former functional, yielding
	\begin{eqnarray}
	D_\alpha(q_\mtn{AB}||q_\mtn{A}\otimes q_\mtn{B})
	&=&
	\frac{\sgn{\alpha}}{\alpha-1}
	\ln\left(
	2m^{1-\alpha}\delta \sum_{i=1}^{m}q_i^{1-\alpha} + (1-2m \delta)^{1-\alpha}\sum_{i=1}^{m}q_i^{1-\alpha}(q_i-2\delta)^\alpha
	\right)
	\end{eqnarray}
	We observe that for all $\alpha\in\mathbb{R}\backslash\{0,1\}$ the former expression does not depend on $n$. Considering now a small $\delta$, we obtain
	\begin{eqnarray}
	\lim_{\delta \to 0^+} D_\alpha(q_\mtn{AB}||q_\mtn{A}\otimes q_\mtn{B})
	\approx
	\frac{\sgn{\alpha}}{\alpha-1}2m\delta \left(m^{-\alpha}\sum_{i=1}^{m}q_i^{1-\alpha}-1\right)
	=
	\frac{\alpha\sgn{\alpha}}{\alpha-1}2m\delta \sum_{i=1}^{m}q_i\left[\frac{\left(\frac{1}{mq_i}\right)^\alpha-1}{\alpha}\right]
	=
	\delta \tau
	\end{eqnarray}
	where $\tau:= 2m\frac{\alpha\,\sgn{\alpha}}{\alpha-1} \sum_{i=1}^{m}q_i \ln_{1-\alpha}\left(\frac{1}{mq_i}\right)$. For $x>0$ and $\alpha\lessgtr 1$ we have, $\ln_{1-\alpha}(x) \lessgtr x-1$ and therefore $\tau>0$ for $\alpha\in\mathbb{R}\backslash\{0\}$. Accordingly, we finally have for small $\delta$,
	\begin{eqnarray}
	I_\alpha(A:B)<\varepsilon:=\delta \tau\,.
	\end{eqnarray}
	The case $\alpha=1$ has been discussed in \cite{mueller2018} and shown to hold as well.

	Regarding the Burg entropy, the entire preceding discussion holds too due to the additivity, subadditivity and strict Schur concavity of the Burg entropy, and the Burg mutual information being equal to
	\begin{eqnarray*}
		I_\mathrm{Burg}(A:B)\equiv D_\mathrm{Burg}(q_\mtn{AB}||q_\mtn{A}\otimes q_\mtn{B})
		=
		-(1-2m\delta)\sum_{i=1}^{m}q_i\ln(q_i-2\delta) + \sum_{i=1}^{m}q_i\ln(q_i) + 2m\delta \ln(m) + (1-2m\delta)\ln(1-2m\delta)
	\end{eqnarray*}
	with
	\begin{eqnarray*}
		\lim_{\delta \to 0^+} D_\mathrm{Burg}(q_\mtn{AB}||q_\mtn{A}\otimes q_\mtn{B})\approx 2m\delta \Bigg(\ln(m)-\sum_{i=1}^{m}q_i\ln(1/q_i)\Bigg)\,.
	\end{eqnarray*}
\end{proof}

\vskip1.0cm

Having proved Theorem \ref{theorem1}, we are now ready to prove Corollary \ref{last3}.

\begin{proof}
	As to the proof, one can follow step by step the proof of Corollary 3 in Ref. \cite{MuellerPastena}, taking into account that we have also proved the subadditivity of the R\'{e}nyi entropy for $\alpha\in(0,+\infty)$ and the Burg entropy which are also continuous, symmetric and additive. Therefore, we define
	\begin{eqnarray}
	I_1(p)=H_\alpha(\eta_m)-H_\alpha(p)\,,\qquad I_2(p)=H_\mathrm{Burg}(\eta_m)-H_\mathrm{Burg}(p)\,,\qquad J(p)=S(\eta_m)-S(p)
	\end{eqnarray}
	and show that $J(p)=\tilde{d}_1\cdot I_1(p)$ and $J(p)=\tilde{d}_2\cdot I_2(p)$, yielding  Eq. (\ref{last2}) with $d_1=\tilde{d}_1/2$, $d_2=\tilde{d}_2/2$ and $c_n=S(\eta_n)+(d_2-d_1)\ln n$.
	In fact, this point whose proof we now provided has already been conjectured in Ref. \cite{MuellerPastena} in the paragraph before the Conclusions.

\end{proof}

\section{Conclusions}\label{SecCon}

We have provided the criteria for the multipartite correlated majorization in terms of the R\'{e}nyi and Burg entropies. The key ingredient has been the subadditivity property of these two entropies, which, to the best of our knowledge, has been provided for the first time in the literature in the present work. The uncorrelated case was already studied, which showed that the necessary and sufficient conditions are given in terms of the R\'{e}nyi and Burg entropies \cite{brandao}. Our result interestingly shows that these two entropies provide the necessary and sufficient conditions also for the correlated case, indicating that correlation does not play a vital role for the multipartite majorization.

Finally, as a corollary of our main result, we show that the unique characterization of the R\'{e}nyi and Burg entropies can be given in terms of continuity, symmetry and (sub)additivity properties. One can compare this result with two other very close approaches. The first one is that of M\"uller and Pastena who showed that the Shannon entropy is the unique entropy that can be characterized by the aforementioned properties \cite{MuellerPastena}. However, the main assumption in \cite{MuellerPastena} is that only the Shannon entropy satisfies the subadditivity property. Once we have shown that both R\'{e}nyi and Burg entropies satisfy the subadditivity, it follows naturally that continuity, symmetry and (sub)additivity properties characterize a linear combination of the R\'{e}nyi and Burg entropies. The second approach is the one adopted by Acz\'{e}l et al. who gave a characterization of the Shannon entropy assuming additivity, symmetry and a weaker form of subadditivity \cite{aczel}. In particular, the weaker form of subadditivity used in \cite{aczel} confines the dimension of the system $B$ to two only (see (ii) in Corollary \ref{last3}). In this work, we considered subadditivity for all dimensions together with the properties of continuity, symmetry and additivity, hence leading to the R\'{e}nyi and Burg entropies as the natural entropies.

\begin{acknowledgments}
The authors, T.O. and G.B.B., gratefully acknowledge Dr. M. M\"uller for the invaluable  correspondence.

\end{acknowledgments}

\appendix

\section{Violation of the subadditivity of the R\'{e}nyi entropy for $\alpha \rightarrow \pm \infty$.}\label{FirstAppendix}

\begin{proof}

	Let us consider the probability distributions $p_\mtn{A}=\{p^A_1, p^A_2, \ldots, p^A_n\}$, $p_\mtn{B}=\{p^B_1, p^B_2, \ldots, p^B_m\}$ and $q_\mtn{A}=\{q^A_1, q^A_2, \ldots, q^A_n\}$, $q_\mtn{B}=\{q^B_1, q^B_2, \ldots, q^B_m\}$. Then,  the R\'{e}nyi infinity  Divergence ($\alpha=+\infty$) for uncorrelated states yields
	\begin{eqnarray}\label{eq01Renyi}
	D_\infty(p_\mtn{A}\otimes p_\mtn{B}||q_\mtn{A}\otimes q_\mtn{B})
	&=&
	\ln\rbr{\max_{ij}R_{ij}}\,,\qquad R_{ij}:= \frac{p^A_ip^B_j}{q^A_iq^B_j}
	\end{eqnarray}
	This means that for the former probabilities, we choose such $i$ and $j$ that $R_{ij}$ takes its maximum value. For clarity, let's call this set of $\{i,j\}$ as $\{i_1,j_1\}$. Then, Eq. (\ref{eq01Renyi}) is written as
	\begin{eqnarray}\label{eq02Renyi}
	D_\infty(p_\mtn{A}\otimes p_\mtn{B}||q_\mtn{A}\otimes q_\mtn{B})
	&=&
	=\ln\rbr{R_{i_1 j_1}}
	=\ln\rbr{ \frac{p^A_{i_1}p^B_{j_1} }{q^A_{i_1}q^B_{j_1} } }
	\end{eqnarray}
	Now, to see the relation between $D_\infty(p_\mtn{A}\otimes p_\mtn{B}||q_\mtn{A}\otimes q_\mtn{B})$ and $D_\infty(p_\mtn{AB}||q_\mtn{A}\otimes q_\mtn{B})$, we rewrite the R\'{e}nyi Divergence in Eq. (\ref{eq02Renyi}) as follows
	\begin{subequations}
		\begin{eqnarray}\label{eq03aRenyi}
		D_\infty(p_\mtn{A}\otimes p_\mtn{B}||q_\mtn{A}\otimes q_\mtn{B})
		&=&
		\ln\rbr{ \frac{p^A_{i_1}p^B_{j_1} }{q^A_{i_1}q^B_{j_1} } }
		=\ln\rbr{ \frac{p^A_{i_1}p^B_{j_1} }{p^{AB}_{i_1j_1} } \cdot \frac{p^{AB}_{i_1j_1} }{q^A_{i_1}q^B_{j_1} }}\\
		\label{eq03bRenyi}
		&=&
		\ln\rbr{ \frac{p^A_{i_1}p^B_{j_1} }{p^{AB}_{i_1j_1} } }
		+
		\ln\rbr{ \frac{p^{AB}_{i_1j_1} }{q^A_{i_1}q^B_{j_1} }}\\
		\label{eq03cRenyi}
		&\leq &
		\ln\rbr{ \frac{p^A_{i_1}p^B_{j_1} }{p^{AB}_{i_1j_1} } }
		+
		D_\infty(p_\mtn{AB}||q_\mtn{A}\otimes q_\mtn{B})
		\end{eqnarray}
	\end{subequations}
	This inequality emerges,  because when the last term in Eq. (\ref{eq03bRenyi}) takes its maximum value, then it is equal to the R\'{e}nyi divergence of the composed system at infinity. Thus,
	\begin{eqnarray}\label{eq04Renyi}
	D_\infty(p_\mtn{AB}||q_\mtn{A}\otimes q_\mtn{B})
	\geq
	D_\infty(p_\mtn{A}\otimes p_\mtn{B}||q_\mtn{A}\otimes q_\mtn{B})
	+
	\ln\rbr{ \frac{p^{AB}_{i_1j_1} }{p^A_{i_1}p^B_{j_1} } }
	\end{eqnarray}
	Here we see that R\'{e}nyi divergence at infinity is superadditive and respectively the R\'{e}nyi entropy is subadditive when $p^{AB}_{i_1j_1} \geq p^A_{i_1}p^B_{j_1}$. However, this depends on the specific case under scrutiny, and cannot be generalized for arbitrary probability vectors. The situation is analogous for $\alpha \rightarrow -\infty$.
	
\end{proof}



\begin{thebibliography}{0}


\bibitem{Cover:2006}
T.~M. Cover and J.~A. Thomas, \textit{Elements of information theory},
2nd~ed.\hskip 1em plus 0.5em minus 0.4em\relax Wiley-Interscience [John Wiley
\& Sons], Hoboken, NJ, 2006.


\bibitem{Shannon_1948}
C.~E. Shannon, ``A mathematical theory of communication,'' \textit{Bell System
	Tech. J.}, vol.~27, pp. 379--423, 623--656, 1948.


\bibitem{Marshall} A. W. Marshall, I. Olkin, and B. C. Arnold, \textit{Inequalities: Theory of Majorization and Its Applications}. New York, NY, USA: Springer, 2010.


\bibitem{horodecki} M. Horodecki and J. Oppenheim, ``Fundamental limitations for quantum
and nanoscale thermodynamics,''\textit{ Nature Commun.}, vol. 4, Jun. 2013, Art. ID 2059.


\bibitem{brandao} F. Brand\~ao, M. Horodecki, N. Ng, J. Oppenheim, and S. Wehner, ``The
second laws of quantum thermodynamics,''\textit{ Proc. Nat. Acad. Sci. USA}, vol. 112, no. 11, pp. 3275-3279, 2015.

\bibitem{trump} D. Jonathan and M. B. Plenio, ``Entanglement-assisted local manipulation of pure quantum states,''\textit{ Phys. Rev. Lett.}, vol. 83, pp. 3566–3569,
Oct. 1999.

\bibitem{Klimesh} M. Klimesh, \textit{Inequalities that Collectively Completely Characterize the Catalytic Majorization Relation}, ArXiv e-prints, September 2007.

\bibitem{Turgut} S. Turgut, ``Necessary and sufficient conditions for the trumping relation,''\textit{ J. Phys. A, Math. Theor.}, vol. 40, 12185–12212, Jul. 2007.
Oct. 1999.


\bibitem{renyi} A. R\'{e}nyi, \textit{On measures of entropy and information}, in: Proceedings of the Fourth Berkeley Symposium on Mathematics, Statistics and Probability, vol. 1, University California Press, Berkeley, 1961, pp. 547-561.


\bibitem{burg} J. P. Burg, \textit{Modern Spectrum Analysis}, D. G. Childers, Ed. New York, NY, USA: IEEE Press, 1978, pp. 34–41.


\bibitem{MuellerPastena} M. P. M\"uller and M. Pastena, \textit{A generalization of majorization that characterizes Shannon entropy}, IEEE Transactions on Information Theory, Vol. 62, No. 4, April 2016


\bibitem{nielsen} M. A. Nielsen and I. L. Chuang, \textit{Quantum Computation and Quantum Information}, Cambridge University Press, 2000.


\bibitem{Nair:2019}
V.~Anantharam, V.~Jog, and C.~Nair, ``{Unifying the Brascamp-Lieb Inequality
	and the Entropy Power Inequality},'' in \textit{2019 IEEE International
	Symposium on Information Theory (ISIT)}, 2019, pp. 1847--1851.

\bibitem{Carlen:2009}
E.~Carlen and D.~Cordero–Erausquin, ``{Subadditivity of The Entropy and its
	Relation to Brascamp–Lieb Type Inequalities},'' \textit{Geometric and
	Functional Analysis}, vol.~19, pp. 373--405, 2009.

\bibitem{Geng:2014}
Y.~Geng and C.~Nair, ``{The capacity region of the two-receiver Gaussian vector
	broadcast channel with private and common messages},'' \textit{IEEE
	Transactions on Information Theory}, vol.~60, no.~4, pp. 2087--2104, April
2014.

\bibitem{Lieb:2002}
E.~H. Lieb, \textit{{Gaussian kernels have only Gaussian maximizers}}.\hskip 1em
plus 0.5em minus 0.4em\relax Springer Berlin Heidelberg, 2002, pp. 595--624.

\bibitem{Carlen:1991}
E.~A. Carlen, ``{Superadditivity of Fisher's information and logarithmic
	Sobolev inequalities},'' \textit{Journal of Functional Analysis}, vol. 101,
no.~1, pp. 194--211, 1991.

\bibitem{Courtade:2014}
T.~A. Courtade and J.~Jiao, ``{An extremal inequality for long Markov
	chains},'' in \textit{2014 52nd Annual Allerton Conference on Communication,
	Control, and Computing (Allerton)}, 2014, pp. 763--770.

\bibitem{Zhang:2018}
X.~Zhang, V.~Anantharam, and Y.~Geng, ``{Gaussian Optimality for Derivatives of
	Differential Entropy Using Linear Matrix Inequalities},'' \textit{Entropy},
vol.~20, no.~3, p. 182, 2018.

\bibitem{Courtade:2018}
T.~A. Courtade, ``{A strong entropy power inequality},'' \textit{IEEE
	Transactions on Information Theory}, vol.~64, no.~4, pp. 2173--2192, April
2018.

\bibitem{Marichal:2000}
J.-L. Marichal, ``{An axiomatic approach of the discrete Choquet integral as a
	tool to aggregate interacting criteria},'' \textit{IEEE Transactions on Fuzzy
	Systems}, vol.~8, no.~6, pp. 800--807, December 2000.

\bibitem{Grabisch:2010}
M.~Grabisch and C.~Labreuche, ``{A decade of application of the Choquet and
	Sugeno integrals in multi-criteria decision aid},'' \textit{Annals of
	Operations Research}, vol. 175, pp. 247--286, 2010.

\bibitem{Duarte:2022}
R.~Pelissari, A. J.~Abackerli and L.~T. Duarte, ``{Choquet capacity identification for multiple criteria sorting problems: A novel proposal based on Stochastic Acceptability Multicriteria Analysis},'' \textit{Applied Soft Computing}, Volume 120, 108727, 2022.

\bibitem{Tsallis2}  C. Tsallis, \textit{Introduction to Nonextensive Statistical Mechanics: Approaching a Complex World,} (\href{https://www.springer.com/gp/book/9780387853581}{Springer, New York, 2009}).


\bibitem{Oik1}
T.~Oikonomou and G.~B. Bagci, ``{Route from discreteness to the continuum for
	the Tsallis $q$-entropy},'' \textit{Physical Review E}, vol.~97, p. 012104,
2018.



\bibitem{Gorban}
A. N. Gorban, P. A. Gorban and G. Judge, ``{Entropy: The Markov Ordering Approach,}'' \textit{Entropy}, vol.~12,  pp. 1145-1193, 2010.


\bibitem{Burg}
J. P. Burg, ``{The relationship between maximum entropy spectra and maximum likelihood spectra,}'' \textit{Geophysics}, vol.~37,  pp. 375-376, 1972.



\bibitem{Perez}
\'{A}.~Capel, A.~Lucia, and D.~P\'{e}rez-Garc\'{\i}a, ``{Superadditivity of
	Quantum Relative Entropy for General States},'' \textit{IEEE Transactions on
	Information Theory}, vol.~64, no.~7, pp. 4758--4765, July 2018.


\bibitem{tsallis_gen}
R.~F. Vigelis, L.~H.~F. de~Andrade, and C.~C. Cavalcante, ``{Properties of a
	Generalized Divergence Related to Tsallis Generalized Divergence},''
\textit{IEEE Transactions om Information Theory}, vol.~66, no.~5, pp. 891--2897, May 2020.

\bibitem{mueller2018}
M. P. M\"uller, ``{Correlating Thermal Machines and the Second Law at the Nanoscale},'' \textit{Physical Review X}, vol.~8, pp. 041051\; 1-23, 2018.



\bibitem{note}
Private communication with Dr. Marcus M\"uller.



\bibitem{aczel}
J. Acz\'el, B. Forte, and C. T. Ng, ``{Why the Shannon and Hartley entropies are ‘natural’,}'' \textit{Adv. Appl. Probab.}, vol.~6, no. 1, pp. 131–146, 1974.

\end{thebibliography}
\end{document}